\documentclass{ws-ijmpa}

\begin{document}

\markboth{Barbieri, Dickhoff}
{Spectroscopic Factors of Asymmetric Isotopes}

%
\catchline{}{}{}{}{}
%

\title{SPECTROSCOPIC FACTORS IN $^{\rm 16}$O AND NUCLEON ASYMMETRY}

\author{C. Barbieri}

\address{Theoretical Nuclear Physics Laboratory, RIKEN Nishina Center,
         2-1 Hirosawa, Wako, Saitama 351-0198 Japan}

\author{W. H. Dickhoff}

\address{Department of Physics, Washington University,
	 St.Louis, Missouri 63130, USA}

\maketitle

\begin{history}
\received{Day Month Year}
\revised{Day Month Year}
\end{history}

\begin{abstract}
The self-consistent Green's functions method is employed to study
the spectroscopic factors of quasiparticle states around $^{\rm 16,28}$O
and $^{\rm 40,60}$Ca. The Faddeev random phase approximation (FRPA) 
is used to account for the coupling of particles with collective excitation 
modes.
Results for ${}^{16}{\rm O}$ are reviewed first.
The same approach is applied to isotopes with large proton-neutron
asymmetry to estimate its effect on spectroscopic factors.
The results, based on the chiral N3LO force, exhibit an asymmetry dependence
similar to that observed in heavy-ion knockout experiments but weaker in 
magnitude.

\end{abstract}


\section{Introduction}
\label{intro}

In the independent particle model (IPM) of atomic nuclei, protons and neutrons
move freely in a common mean-field potential. Obviously, this is an approximate
picture since the residual interaction smears the Fermi surface and leads
to partial occupation of each orbit.
For states close to the Fermi level, this effect is observed through
a reduction of the experimental knockout and pickup cross sections.
Experimental spectroscopic factors (SFs) are defined as the quenching of
the observed reaction rate with respect to that calculated assuming
full occupancy. Hence, they are interpreted as the occupation of a given orbit.
However, from a strict theoretical point of view SFs are not occupation numbers.
Instead, they give a ``measure'' of what fraction of the {\em final} wave function
can be factorized into a (correlated) core plus an independent particle
or hole state.
Strong deviations from unity signal the onset of substantial correlation effects
and imply the existence of non trivial many-body dynamics. 

As far as stable nuclei are concerned, a large body of data has been 
accumulated from ($e$,$e'p$) experiments yielding the best estimates of 
absolute spectroscopic factors. 
These studies showed that proton SFs for isotopes 
all across the nuclear chart are uniformly quenched to 60-70\%
of the IPM value~\cite{lap93,kramers01}.
This information is however limited to protons and stable isotopes.
More recently, experimental information on SFs of dripline isotopes has been
obtained by means of nucleon knockout using intermediate energy heavy-ion
beams~\cite{Han.03,Gad.08}. These results also include neutron data, and 
suggest a strong dependence of SFs on the N/Z ratio.

In nuclear structure studies, one usually distinguishes
between short-range (SRC) and long-range and correlations (LRC).
The former are induced by the strong repulsive core and the tensor 
component of the nuclear force at small distances.
 These remove strength from the Fermi sea to very high energies and momenta and
have long been proposed as one possible mechanism for the quenching
of SFs.
 However, different theoretical evaluations predicted that SRC can account
for at most a 10-15\% reduction (see Refs.~\cite{DMP,O16com} for $^{16}$O).
This result is now supported by electron scattering
experiments~\cite{danielaPRL} at high energies, where the reactions can be
analyzed using a Glauber-inspired approach~\cite{Bar.04b,Bar.05,Bar.06a}
(see also Ref.~\cite{Science}).
 Calculations like those reported in Sec.~\ref{sec:3} shows that
the largest part of the quenching is instead due to LRC. In particular,
couplings between single nucleons and collective surface phonons
is important~\cite{DiB.04}.

This talk reviews our theoretical understanding of SFs obtained 
from calculations based on Green's function theory.
The formalism used in these studies is summarized
in Sec.~\ref{sec:2} and its application to $^{16}$O, as a test case,
is discussed in Sec.~\ref{sec:3}.
Sec.~\ref{sec:4} reports on a first investigation of correlations
in asymmetric isotopes.

\section{Faddeev-RPA Method}
\label{sec:2}

\begin{figure}[t]
 \begin{center}
 \includegraphics[height=1.45in]{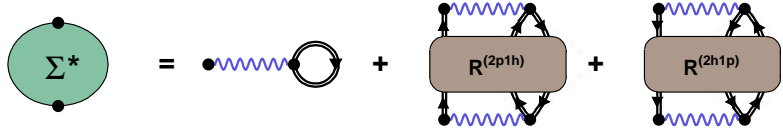}
 \end{center}
\vspace{-0.5in}
\caption[]{
 Diagrams contributing to the irreducible self-energy $\Sigma^\star(\omega)$.
   The double lines represent a dressed propagator and 
   the wavy lines correspond to a G-matrix (that is used in this work as an 
   effective interaction).  The first term is the 
   Brueckner-Hartree-Fock potential while the others represent 
two-particle--one-hole / two-hole--one-particle (2p1h/2h1p) 
   or higher contributions that are approximated through the Faddeev RPA
   equations.
\label{fig:selfenergy} }
\end{figure}

We consider the calculation of the single-particle (sp) Green's function~\cite{DiVbook}
\begin{equation}
 g_{\alpha \beta}(\omega) ~=~ 
 \sum_n  \frac{ \left( {\cal X}^{n}_{\alpha} \right)^* \;{\cal X}^{n}_{\beta} }
                       {\omega - (E^{A+1}_n - E^A_0) + i \eta }  ~+~
 \sum_k \frac{ {\cal Y}^{k}_{\alpha} \; \left( {\cal Y}^{k}_{\beta} \right)^* }
                       {\omega - (E^A_0 - E^{A-1}_k) - i \eta } \; ,
\label{eq:g1}
\end{equation}
where ${\cal X}^{n}_{\alpha} = {\mbox{$\langle {\Psi^{A+1}_n} \vert $}}
 c^{\dag}_\alpha {\mbox{$\vert {\Psi^A_0} \rangle$}}$%
~(${\cal Y}^{k}_{\alpha} = {\mbox{$\langle {\Psi^{A-1}_k} \vert $}}
 c_\alpha {\mbox{$\vert {\Psi^A_0} \rangle$}}$) are the
spectroscopic amplitudes for the excited states of a system with
$A+1$~($A-1$) particles.
In these definitions, $\vert\Psi^{A+1}_n\rangle$, $\vert\Psi^{A-1}_k\rangle$ 
are the eigenstates, and $E^{A+1}_n$, $E^{A-1}_k$ the eigenenergies of the 
($A\pm1$)-nucleon system. Therefore, the poles of the sp propagator reflect
the energy transfer observed in pickup and knockout reactions.
 The corresponding SFs for transitions to a quasihole~(quasiparticle) state
are obtained as
$Z_k = \sum_{\alpha} \left| {\cal Y}^{k}_{\alpha} \right|^2$~%
\hbox{($Z_n = \sum_{\alpha} \left| {\cal X}^{n}_{\alpha} \right|^2$)}.

The one-body Green's function is computed by solving the Dyson equation
\begin{equation}
 g_{\alpha \beta}(\omega) =  g^{0}_{\alpha \beta}(\omega) \; +  \;
   \sum_{\gamma \delta}  g^{0}_{\alpha \gamma}(\omega) 
     \Sigma^\star_{\gamma \delta}(\omega)   g_{\delta \beta}(\omega) \; \; ,
\label{eq:Dys}
\end{equation}
where the irreducible self-energy $\Sigma^\star_{\gamma \delta}(\omega)$ acts
as an effective, energy-dependent, potential. The latter can be expanded in a
Feynman-Dyson series~\cite{DiVbook} in terms of the exact
propagator $g_{\alpha \beta}(\omega)$, which itself is a solution
of Eq.~(\ref{eq:Dys}).
 In this expansion, $\Sigma^\star_{\gamma  \delta}(\omega)$ can be represented
as shown in Fig.~\ref{fig:selfenergy} by the sum
of a Hartree-Fock-like potential and the polarization propagator, $R(\omega)$,
that account for deviations from the mean-field~\cite{DiB.04}. It is at
the 2p1h/2h1p level that the correlations involving couplings of sp
to collective modes need to be included.
 The SCGF approach is initiated by solving the self-energy
and the Dyson Eq.~(\ref{eq:Dys}) in terms of an unperturbed
propagator (e.g. Hartree-Fock). The (dressed) solution 
$g_{\alpha \beta}(\omega)$ is then taken as a new input and
the whole calculation is iterated until convergence is reached.

\subsection{Faddeev RPA Method for the Self-Energy}
\label{sec:2fadd}

The polarization propagator $R(\omega)$ can be expanded in terms of simpler
Green's functions that involve the propagation of one quasiparticle
[Eq.~(\ref{eq:g1})] or more. 
This approach has the advantage that it allows the identification and 
inclusion of key physics ingredients of the many-body dynamics. 
By truncating to particular
subsets of diagrams, one can then construct suitable approximations to
the self-energy. Moreover, since infinite sets of linked diagrams are summed, 
the approach is non-perturbative and satisfies the extensivity condition. This
expansion also serves as a guideline for systematic improvements of the method.

 In the following we are interested in describing the coupling of
sp motion to particle-hole (ph) and particle-particle (hole-hole) [pp(hh)] 
collective excitations of the system.
Following Refs.~\cite{Bar.01,Bar.07}, we first consider the ph
polarization propagator describing excited states in the $A$-particle system 
\begin{eqnarray}
 \Pi_{\alpha \beta , \gamma \delta}(\omega) &=& 
 \sum_{n \ne 0}  \frac{  {\mbox{$\langle {\Psi^A_0} \vert $}}
            c^{\dag}_\beta c_\alpha {\mbox{$\vert {\Psi^A_n} \rangle$}} \;
             {\mbox{$\langle {\Psi^A_n} \vert $}}
            c^{\dag}_\gamma c_\delta {\mbox{$\vert {\Psi^A_0} \rangle$}} }
            {\omega - \left( E^A_n - E^A_0 \right) + i \eta } 
\nonumber \\
 &-& \sum_{n \ne 0} \frac{  {\mbox{$\langle {\Psi^A_0} \vert $}}
              c^{\dag}_\gamma c_\delta {\mbox{$\vert {\Psi^A_n} \rangle$}} \;
                 {\mbox{$\langle {\Psi^A_n} \vert $}}
             c^{\dag}_\beta c_\alpha {\mbox{$\vert {\Psi^A_0} \rangle$}} }
            {\omega - \left( E^A_0 - E^A_n \right) - i \eta } \; ,
\label{eq:Pi}
\end{eqnarray}
and the two-particle propagator, for the addition/removal of two 
particles
\begin{eqnarray}
 g^{II}_{\alpha \beta , \gamma \delta}(\omega) &=& 
 \sum_n  \frac{  {\mbox{$\langle {\Psi^A_0} \vert $}}
                c_\beta c_\alpha {\mbox{$\vert {\Psi^{A+2}_n} \rangle$}} \;
                 {\mbox{$\langle {\Psi^{A+2}_n} \vert $}}
         c^{\dag}_\gamma c^{\dag}_\delta {\mbox{$\vert {\Psi^A_0} \rangle$}} }
            {\omega - \left( E^{A+2}_n - E^A_0 \right) + i \eta }
\nonumber \\  
&-& \sum_k  \frac{  {\mbox{$\langle {\Psi^A_0} \vert $}}
    c^{\dag}_\gamma c^{\dag}_\delta {\mbox{$\vert {\Psi^{A-2}_k} \rangle$}} \;
                 {\mbox{$\langle {\Psi^{A-2}_k} \vert $}}
                  c_\beta c_\alpha {\mbox{$\vert {\Psi^A_0} \rangle$}} }
            {\omega - \left( E^A_0 - E^{A-2}_k \right) - i \eta } \; .
\label{eq:g2}
\end{eqnarray}
 In the calculation of Sec.~\ref{sec:3}, these
have been approximated by solving the dressed random phase
approximation (DRPA)
equations~\cite{DiVbook}, which account for the effects of
the strength distribution of the particle and hole fragments.

\begin{figure}
 \begin{center}
    \parbox[b]{.25\linewidth}{
\hspace{.2in} \includegraphics[height=2.2in]{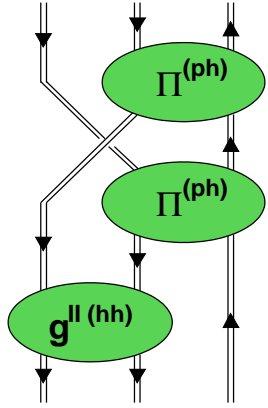} }
    \hfill
    \parbox[b]{.69\linewidth}{
\caption[]{Example of a diagram appearing in the all-order summation
generated by the set of Faddeev equations. This diagram contributes to
the $R^{2h1p}(\omega)$ propagator seen in Fig.~\ref{fig:selfenergy}.
 \vspace{.4in} 
\label{fig:FaddSum} }  } 
\end{center}
\end{figure}

The ph~(\ref{eq:Pi}) and pp(hh)~(\ref{eq:g2}) propagators are inserted
in the self-energy by solving a set of Faddeev equations
for the 2p1h and 2h1p propagators of Fig.~\ref{fig:selfenergy}.
The details of the Faddeev RPA (FRPA) approach are given
in Ref.~\cite{Bar.01,Bar.07}.
For the present discussion it is sufficient to note that collective
excitations are coupled to sp propagators generating an infinite series
of diagrams, including the one shown in Fig.~\ref{fig:FaddSum}.

\section{Hole Spectroscopic Factors for $^{16}$O}
\label{sec:3}

 The calculations for $^{16}$O, in  Ref.~\cite{Bar.02}, were performed in
an harmonic oscillator basis with parameter $b =$ 1.76~fm (corresponding to
$\hbar\omega =$ 13.4~MeV). The first four major
shells (from $0s$ to $1p0f$) plus the $0g_{9/2}$ orbit where included.
 Inside the model space, a Brueckner G-matrix derived from
the Bonn-C potential
 was used as an effective interaction.
The short-range core of this NN interaction induces an additional 10\%
reduction of the spectroscopic factors for main quasiparticle peaks,
by moving strength to very high energies~\cite{DiB.04}.
This effect was included in the solution of the Dyson equation by treating
the energy dependence of the G-matrix explicitly.

The FRPA calculations were then iterated to self-consistency, thus
including the effects of fragmentation.
In doing this, the largest fragments that appear---close to the Fermi
energy---in the (dressed) sp propagator, Eq.~(\ref{eq:g1}), are
maintained. The remaining strength is collected, at each iteration, into
effective poles~\cite{Bar.02}.

\subsection{Spectroscopic Factors and Role of Low Excited States in ${}^{16}{\rm O}$}
\label{sec:3:1}

\begin{figure}
 \begin{center}
 \includegraphics[height=2.9in]{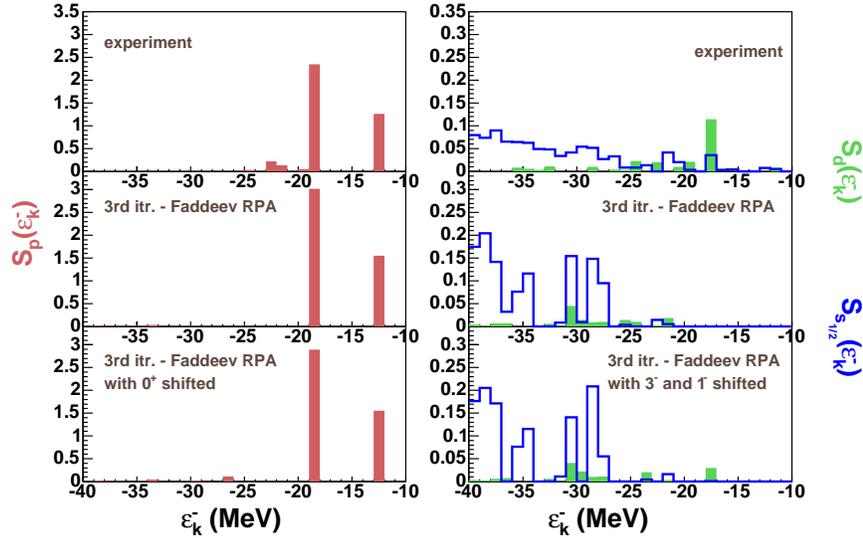}
 \end{center}
\caption[]{One-proton removal strength as a function of the 
   hole sp energy $\varepsilon^{-}_k = E^A_0 - E^{A-1}_k$
    for ${}^{16}{\rm O}$ for angular momenta
  $\ell =1$ (left) and $\ell =0,2$ (right).
   For the positive parity
   states, the solid bars correspond
   to results for $d_{5/2}$ and $d_{3/2}$ orbitals, while the
   thick lines refer to $s_{1/2}$.  
   The top panels show the experimental values taken from~\cite{leus}.
   The central panels give the theoretical results for the self-consistent
   spectral function.
   The bottom panels show the results obtained by constraining the lowest
   excited states [i.e. the poles of Eq.~(\ref{eq:Pi})], to the corresponding
   experimental energies.
   A small amount of $p_{3/2}$ strength is observed experimentally at around 
   -23~MeV (top panel) and obtained theoretically at -26~MeV (bottom panel).
\label{fig:p+sd_hsf} }
\end{figure}

The one-hole strength distribution obtained upon convergence of the 
SCGF is shown in Fig.~\ref{fig:p+sd_hsf}, where it is compared with the
experiment (top panels).
Similar results are obtained for the particle strength, including large peaks
near the Fermi surface and a fragmented distribution at larger energies. 
The spectroscopic factors obtained for removal of a proton amount
to 0.75 of the IPM value for the $p_{3/2}$ peak and 0.77 for the $p_{1/2}$.
These refer to the middle panels of this figure.

Failures to fully reproduce the observed removal strength can be 
traced to the difficulties of the RPA to describe the low-energy 
spectrum. 
 To clarify this point we repeated the above calculations of the
sp propagator by shifting, at each iteration, the first RPA $0^+$
eigenstate in ${}^{16}{\rm O}$.  This eigenvalue is obtained at larger
energies but the corresponding pole of Eq.~(\ref{eq:Pi}) was constrained
to the experimental energy of the first excited state.
 This is expected to represent qualitatively
the cluster state which is also excited by a M(E0) transition (see e.g.
the contribution of Funaki and Schuck to these proceedings).%
\footnote{The RPA wave function is a particle-hole configuration
and is intrinsically different from the experimental first excited
level, which is a cluster state.
However, both states are strongly excited by one-body operators. Since
this is the mechanism that couples nucleons to collective modes in the
FRPA, Fig.~\ref{fig:FaddSum}, the two are expected to have similar effects
on the results discussed here.}
 This change leads to the appearance of satellite $p_{3/2}$ fragments around
-26.3~MeV, which might be identified with the fragments seen experimentally
at slightly higher energy. These are calculated to have total strength
of 2.6\%, while the dominant peak is reduced to 0.72 (see Table~\ref{tab:SpectFact}).
 The associated spectral function is shown in the lower-left panel
of Fig.~\ref{fig:p+sd_hsf}.
Thus one concludes that the additional $p_{3/2}$ fragments are generated
by propagating a hole on top of excited $0^+$ level of the
${}^{16}{\rm O}$ core.

The other two low-lying states of ${}^{16}{\rm O}$ that may be of some
relevance are the isoscalar $1^-$ and $3^-$, and are calculated at
$\sim$3~MeV above the experimental value. Constraining their energies to the 
experimental values leads to analogous improvements.
In particular a  $d_{5/2}$ hole peak is obtained at a missing energy
of -17.7~MeV, in nice agreement with the experiment%
~(lower-right panel in Fig.~\ref{fig:p+sd_hsf})%
.

\subsection{Status of theoretical calculations for $^{16}O$}

 \begin{table}
\tbl{Dependence of theoretical spectroscopic factors (as a fraction of
    the IPM value) on the inclusion of various SRC and LRC effects.
     The value in parentheses is the summed
    strength of small $p_{3/2}$ fragments near the quasihole peak.}
   { \begin{tabular}{lrcrc}
    \hline
         {Spectroscopic factors of $^{16}$O}
       & \hspace{1.0cm}
       & {$Z_{p1/2}$}
       & \hspace{0.5cm}
       & {$Z_{p3/2}$}   \\
    \hline
    SRC only~\cite{DMP,O16com}                                   &  & $\sim$0.90 &  & $\sim$0.90 \\
    SRC + LRC(FRPA + sp dressing)~\cite{Bar.02}                  &  &    0.77 &  &       0.75 \\
    SRC + LRC(FRPA + sp dressing + 0$^+_2$ state)~\cite{Bar.02}  &  &    0.77 &  &       0.72(0.026) \\
    \\
    experiment~\cite{radici,Meucci} &  & 0.64-0.71 &  & 0.54-0.61 \\
    \hline
    \end{tabular}
    \label{tab:SpectFact} }
 \end{table}

The contributions to the reduction of SFs are summarized in 
Table~\ref{tab:SpectFact} for the $p$ shell orbits of $^{16}$O. The range
of values labeled as ``experiment''  reflects the quenching factors {\em needed}
by theoretical analyses to reproduce the observed ($e$,$e'p$) data.
As mentioned in Sec.~\ref{intro}, SRC account for a modest part of
the reduction of SFs. For the case of hole states in $^{16}$O,
both Green's function theory~\cite{DMP} and
variational Monte Carlo~\cite{O16com} methods predict a 10\% depletion.
 The agreement between different methods and interactions (Bonn and Argonne
forces were used) give us confidence that short-range physics is
under control and that, at low energies, its overall effect exhibits little
dependence on
how realistic forces are modeled. 
 The calculations of Sec.~\ref{sec:3:1} shows that the largest part of the
quenching is instead due to LRC effects, such as couplings between single
nucleons and collective surface phonons and configuration mixing in the
valence shell. 
 Based on the findings of Ref.~\cite{Bar.02}, it is plausible that clustering degrees
of freedom are the missing ingredient in this particular isotope.
We note that $^{16}$O is an exceptionally difficult nucleus in this respect
and so far it has defeated a complete theoretical understanding of knockout
cross sections.
Thus it provides a stringent benchmark for theoretical calculations.
Successful calculations have been obtained for other nuclei~\cite{ca48,LapLi7}.

\section{Extension to Larger Systems and Asymmetric Nuclei}
\label{sec:4}

Experimental and theoretical studies of electron scattering reactions have led
to a global picture of the properties of protons for stable 
nuclei~\cite{DiB.04}.
The first information on how these features change toward the driplines has
only recently become available. In particular, one-nucleon knockout
experiments in inverse kinematics have found that SFs do change
with proton-neutron asymmetry.
In general, the quenching of quasiparticle orbits (and hence
correlations) become stronger with increasing separation energy~\cite{Gad.08}.

The extension of FRPA calculations from $^{16}$O to large isotopes faces
technical challenges due to the increasing model space and computational load.
A first breakthrough has been obtained in Ref.~\cite{Bar.06}, where 
self-consistent calculations of $^{16}$O were obtained in a large basis 
(up to 8 oscillator shells). Similar 
calculations can now be performed in the $pf$ shell. 
In this case self-consistency
is implemented only partially (for now), according to Ref.~\cite{BaH.09}.

\begin{figure}[pb]
\centerline{\psfig{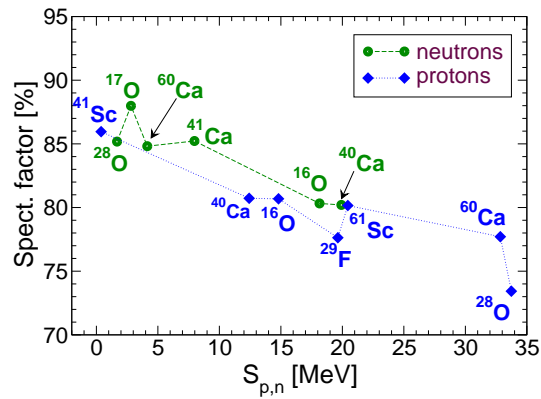}}
\vspace*{8pt}
\caption{SFs obtained from partially self-consistent FRPA. All numbers are 
  given as a fraction of the IPM value and refer to transitions from ground
  state to ground state. The points refer to knockout of a nucleon from the 
  isotope indicated nearby. The lines are a guide to the eye. \label{SFasymm}}
\end{figure}

Figure~\ref{SFasymm} shows first FRPA results for the spectroscopic factors
of quasiparticles around $^{\rm 16,28}$O and $^{\rm 40,60}$Ca. These
are obtained form a G-matrix~\cite{gmtx} based on the chiral N3LO interaction
and following the calculation scheme of Ref.~\cite{BaH.09}.
A dependence on the proton-neutron asymmetry is indeed observed in the FRPA,
with the spectroscopic factors becoming smaller with increasing nucleon
separation energy.  A dispersive optical model analysis, which is constrained
to data up to $^{\rm 48}$Ca, has also been extrapolated to proton rich Ca
isotopes, with similar findings~\cite{Cha.06,Cha.07}.
 However, for both analyses the change in magnitude is significantly 
smaller than the
one deduced from direct knockout data~\cite{Gad.08}. 
The most asymmetric isotope in Fig.~\ref{SFasymm} is $^{28}$O with an
asymmetry parameter $\alpha=(N-Z)/A\approx0.43$. 
In this case, the FRPA SFs for knockout of a proton and a neutron differ by 
about 11\% of the IPM.
This result is in agreement with calculations of nuclear matter where, however,
the only mechanism considered was SRC~\cite{Fri.05}.
It must be stressed that collective excitations are the most important
degrees of freedom governing the reduction of SFs. These are properly
accounted for by the FRPA approach. However, realistic two-nucleon forces
such as the one used here have a tendency to overestimate the excitation
energy of giant resonances, and therefore to underestimate their importance.
It is plausible that the dependence on asymmetry seen in Fig.~\ref{SFasymm}
will become more substantial once FRPA calculations with improved forces will
be available.

\section{Conclusions}
\label{sec:concl}

The Faddeev RPA method has been used to investigate the effects of SRC and 
couplings of nucleons to collective excitations. The latter give the most
substantial contribution to the quenching of spectroscopic factors.
Applications to asymmetric isotopes shows a reduction of the SFs for
quasiparticle orbits with increasing separation energy.
This dependence is considerably weaker than the one deduced from heavy-ion 
knockout experiments.
 It is argued that improved forces and calculations will be needed
to resolve this issue.

\section*{Acknowledgments}

This work was in part supported by the U.S. National Science Fundation
under Grant no. PHY-0652900.
\appendix


\begin{thebibliography}{00}  

\bibitem{lap93}
L.~Lapik\'{a}s, 
\newblock Nucl. Phys. {\bf A553}, 297c (1993).

\bibitem{kramers01}
G.~J. Kramers, H.~P. Blok, and L. Lapik\'as,
\newblock Nucl. Phys. {\bf A679}, 267 (2001).


\bibitem{Han.03}
P.~G.~Hansen and J.~A.~Tostevin,
\newblock Ann. Rev. Nucl. Part. Sci. {\bf 53}, 219 (2003).


\bibitem{Gad.08}
A.~Gade, {\em et al.},
\newblock Phys.\ Rev.\ C {\bf 77}, 0044306 (2008).


\bibitem{DMP}
H. M{\"{u}}ther, A. Polls, and W. H. Dickhoff,
\newblock Phys. Rev. C \textbf{51}, 3040 (1995).


\bibitem{O16com}
D.~Van~Neck,{\em et al}.,
\newblock Phys. Rev. C {\bf 57}, 2308 (1998).




\bibitem{danielaPRL}
D.~Rohe, {\em et al}.,
\newblock Phys. Rev. Lett. {\bf 93}, 182501 (2004).


\bibitem{Bar.04b}
C.~Barbieri  and L.~Lapik\'as,
\newblock Phys.\ Rev.\ C {\bf 70}, 054612 (2004).


\bibitem{Bar.05}
C.~Barbieri, D.~Rohe, I. Sick, and L.~Lapik\'as,
\newblock Phys.\ Lett.\ {\bf B608}, 47 (2005).


\bibitem{Bar.06a}
C.~Barbieri,
\newblock Nucl.\ Phys.\ B\ (Proc.\ Suppl.) {\bf 159}, 174 (2006).


\bibitem{Science}
R.~Subedi {\em et al.},
\newblock Science {\bf 320}, 1476 (2008).


\bibitem{DiB.04}
W.~H.~Dickhoff and C.~Barbieri,
\newblock Prog. Part. Nucl. Phys. {\bf 52}, 377 (2004).

\bibitem{DiVbook} W.~H.~Dickhoff and D.~Van~Neck,
{\em Many-Body Theory Exposed!}, 2nd ed. (World Scientific, Singapore, 2008).


\bibitem{Bar.01}
C.~Barbieri and W.~H.~Dickhoff,
\newblock Phys.\ Rev.\ C {\bf 63}, 034313 (2001).


\bibitem{Bar.07}
C.~Barbieri, D. Van Neck and W.~H.~Dickhoff,
\newblock Phys.\ Rev.\ A {\bf 76}, 052503 (2007).


\bibitem{Bar.02}
C.~Barbieri and W.~H.~Dickhoff,
\newblock Phys.\ Rev.\ C {\bf 65}, 064313 (2002).


\bibitem{leus}
M.~Leuschner {\em et al.},
\newblock Phys. Rev. C {\bf 49}, 955 (1994).


\bibitem{radici}
M. Radici, W. H. Dickhoff, and E. Roth Stoddard,
Phys. Rev. C \textbf{66}, 014613 (2002).


\bibitem{Meucci}
M.~Radici, A.~Meucci and W.~H.~Dickhoff,
\newblock Eur.\ Phys.\ J.\ {\bf A 17}, 65 (2003).


\bibitem{ca48}
G.~A.Rijsdijk, K~Allaart, and W.~H.~Dickhoff,
\newblock Nucl. Phys. {\bf A550}, 159 (1992).


\bibitem{LapLi7}
L.~Lapik\'{a}s, J.~Wesseling, and R.~B.~Wiringa,
\newblock Phys. Rev. Lett. {\bf 82}, 4404 (1999).


\bibitem{Bar.06}
C.~Barbieri,
\newblock Phys.\ Lett.\ {\bf B643}, 268 (2006).


\bibitem{BaH.09}
C.~Barbieri and M. Hjorth-Jensen,
\newblock in preparation.


\bibitem{gmtx}
M.~Hjorth-Jensen, T.~T.~S.~Kuo, and E.~Osnes,
\newblock Phys. Rep. {\bf 261}, 125 (1995).


\bibitem{Cha.06}
R.~J.~Charity {\em et al.},
\newblock Phys.\ Rev.\ Lett. {\bf 97}, 162503 (2006).


\bibitem{Cha.07}
R.~J.~Charity {\em et al.},
\newblock Phys.\ Rev.\ C {\bf 76}, 044314 (2007).


\bibitem{Fri.05}
T.~Frick {\em et al.},
\newblock Phys.\ Rev.\ C {\bf 71}, 014313 (2005).



\end{thebibliography}
\end{document}